\documentclass[10pt,
							 twocolumn,
							 superscriptaddress,
							 english,
							 prl,
							 showpacs,
							 floatfix,
							 aps
							]{revtex4-1}
\usepackage[utf8]{inputenc}
\usepackage{amsmath}
\usepackage{amssymb}
\usepackage{graphicx}
\usepackage{xspace}
\usepackage{ulem}

\makeatletter

\newcommand*{\didv}{\ensuremath{\mathrm{d}I/\mathrm{d}V}\xspace}
\newcommand*{\Figref}[1]{Figure~\ref{#1}}
\newcommand*{\mos}{MoS$_2$}

\usepackage{soul}
\usepackage{braket}
\usepackage[backref=none,
bookmarksnumbered=true,
bookmarks=true,
bookmarksopen=true,
colorlinks=true,
citecolor=blue,
linkcolor=blue,
anchorcolor=green,
urlcolor=blue,unicode=false]{hyperref}

\usepackage{babel}

\begin{document}

\title{Monolayers of \mos\ on Ag(111) as decoupling layers for organic molecules: resolution of electronic and vibronic states of TCNQ}
\author{Asieh Yousofnejad}
\author{Ga\"el Reecht}
\author{Nils Krane}
\author{Christian Lotze}
\author{Katharina J. Franke}

\affiliation{Fachbereich Physik, Freie Universit\"at Berlin, Arnimallee 14, 14195 Berlin, Germany}

\begin{abstract}
The electronic structure of molecules on metal surfaces is largely determined by hybridization and screening by the substrate electrons. As a result, the energy levels are significantly broadened and molecular properties, such as vibrations are hidden within the spectral lineshapes. Insertion of thin decoupling layers reduces the linewidths and may give access to the resolution of electronic and vibronic states of an almost isolated molecule. Here, we use scanning tunneling microscopy and spectroscopy to show that a single layer of \mos\ on Ag(111) provides a semiconducting band gap that may prevent molecular states from strong interactions with the metal substrate. We show that the lowest unoccupied molecular orbital (LUMO) of tetra-cyano-quino-dimethane (TCNQ) molecules is significantly narrower than on the bare substrate and that it is accompanied by a characteristic satellite structure. Employing simple calculations within the Franck-Condon model, we reveal their vibronic origin and identify the modes with strong electron-phonon coupling.
\end{abstract}
\maketitle

\section{Introduction}

When molecules are adsorbed on metal surfaces, their electronic states are strongly perturbed by hybridization, charge transfer and screening \cite{Lu2004, Thygesen2009, Braun2009,Tautz2007}. These effects lead to a broadening and shifting of the molecular resonances \cite{Torrente2008}. Often the molecular functionality is also lost due to these interactions \cite{Tegeder2012}. However, addressing individual molecules in devices or by single-molecule spectroscopy as offered in a scanning tunneling microscope, requires a metal electrode. To (partially) preserve the molecular properties the molecule--electrode coupling has to be properly designed. An elegant way is to clamp the molecule between electrodes via weak single-atom bonds at opposing sites of the molecule while the molecule is freely hanging between the electrodes \cite{Reed1997, Reichert2002, Venkatarman2006, Lafferentz2009}. While these configurations give access to important transport properties \cite{Nitzan2003,Tao2006, Aradhya2013}, they do not allow for imaging molecular properties with intramolecular resolution \cite{Reecht2016}. The latter requires the molecules to be flat lying on a surface. To decouple such flat lying molecules from a metal, thin insulating layers have been engineered, ranging from ionic salts \cite{Repp2005a,Liljeroth2007}, over oxides \cite{Qiu2003, Heinrich2004, Rau2014}, nitrides \cite{Hirjibehedin2006}, and molecular layers \cite{Franke2008, Matino2011} to 2D materials, such as graphene \cite{Garnica2013,Riss2014}, and hexagonal boron nitride \cite{Schulz2013}. 

The most recent development of decoupling layers made use of the in-situ fabrication of single-layers of transition-metal-dichalcogenides on metal surfaces. A monolayer of \mos\ on Au(111) provided very narrow molecular resonances, close to the thermal resolution limit at 4.6\,K \cite{Krane2018}. The exquisite decoupling efficiency has been ascribed to a combination of its rather large thickness of three atomic layers, its electronic band gap, and its non-ionic nature. All together, these properties prohibited fast electronic relaxations into the metal and coupling to phonons, which otherwise led to lifetime broadening \cite{Repp2005, Fatayer2018}. 

The electronic properties of \mos\ on a metal surface are not the same as of a free-standing monolayer. Both theory and experiment have found considerable hybridization of electronic states at the interface \cite{Bruix2016}. As a consequence, the band gap is narrowed. Instead of the predicted band gap of 2.8\,eV of the free-standing layer \cite{Cheiwchanchamnangij2012, Qiu2015}, the band gap of the hybrid structure amounts to only $\sim 1.7$\,eV \cite{Bruix2016}. Interestingly, the states at the $K$ point are much less affected than the states at $\Gamma$. Hence, the system remains promising for optoelectronic devices with selective access to the spin-orbit-split bands at $K$ and $K'$ by circularly polarized light \cite{Bana2018}. 

The potential as decoupling layer for molecules, may become even more appealing by the fact that monolayers of transition-metal-dichalcogenides can be grown in-situ on different metal surfaces, where the precise hybridization and band alignment depends on the nature of the substrate \cite{Dendzik2017}. One may thus envision tuning the band gap alignment for decoupling either the molecules' lowest unoccupied (LUMO) or highest occupied molecular orbitals (HOMO). 

While \mos\ on Au(111) has already been established as an outstanding decoupling layer \cite{Krane2018}, we will now explore this potential for \mos\ on a Ag(111) surface. In agreement with the band modifications of WS$_2$ on Au(111) and Ag(111), we find that the band gap remains almost the same, but shifted to lower energies \cite{Dendzik2017}. As a test molecule we chose tetra-cyano-quino-dimethane (TCNQ). Due to its electron-accepting character, this choice will allow us to detect a negative ion resonance within the band gap of \mos. We will show that the LUMO is indeed decoupled from the metallic substrate as we can detect a narrow linewidth followed by a satellite structure. We can reproduce this fine structure by simulating the vibronic states of the gas-phase molecule.


\section{Results and Discussion}
We have grown monolayer islands of \mos\ on an atomically clean Ag(111) surface, which had been exposed to sputtering-annealing cycles under ultrahigh vacuum previously. The growth procedure was adapted from the case of of \mos\ on Au(111) \cite{Gronborg2015,Krane2016}, with Mo deposition on the surface in an  H$_2$S atmosphere of $5\cdot 10^{-5}$ mbar, while the sample is annealed to 800\,K.
Tetra-cyano-quino-dimethane (TCNQ) molecules were deposited on the as-prepared sample held at 230\,K. The sample was then cooled down and transferred to the scanning tunneling microscope (STM). All measurements were performed at 4.6\,K. Differential conductance (\didv) maps and spectra were taken with a lock-in amplifier at modulation frequencies of 812-921\,Hz, with the amplitudes given in the figure captions.

\begin{figure*}[h!]
\includegraphics[width=16cm]{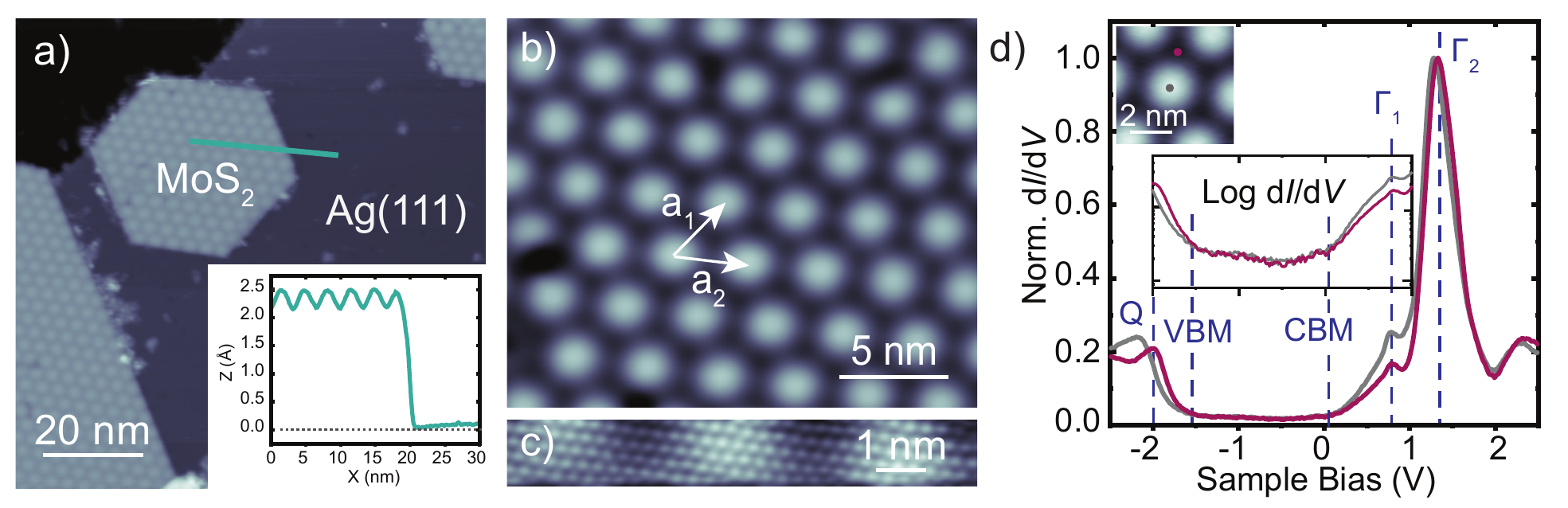}
\caption{a) STM topography of MoS$_2$ on Ag(111) recorded at V = 1.2\,V, I = 20\,pA. Inset: Line profile of a monolayer \mos\ island along the green line. b) Close-up view on the moir\'e structure. c) Atomically resolved terminating S layer (V = 5\,mV, I = 1\,nA).
d) Constant-height \didv spectra on the MoS$_2$/Ag(111) recorded on top and on hollow region of the moir\'e structure as shown on the inserted STM topography (feedback opened at V = 2.5\,V, I = 0.5\,nA, V$_\mathrm{mod}=10$\,mV). The inset shows the gap region of MoS$_2$/Ag(111) in logarithmic scale. We identify the VBM and CBM as the change in slope of the \didv signal. Dashed lines indicate the conduction band minimum (CBM) at $\sim 0.05$\,V and valence band maximum (VBM) at $\sim -1.55$\,V. The strong features in the \didv spectra are associated to the onset of specific bands, which are labeled by $Q$, $\Gamma_1$ and $\Gamma_2$ according to their location in the Brillouin zone. The assignment follows Ref. \cite{Krane2018a}.}
\label{fig1}

\end{figure*}

\subsection{Characterization of single-layer \mos\ on Ag(111)}

\Figref{fig1}a presents an STM image of the Ag(111) surface after the growth of \mos\ as described above. We observe islands with tens to hundreds of nanometer diameter and of $2.3 \pm 0.2$\, \AA\ apparent height (inset of \Figref{fig1}). The apparent height is much smaller than the layer distance in bulk \mos\ \cite{Wakabayashi1975} due to electronic-structure effects, but in agreement with a single layer of \mos\ on a metal surface \cite{Gronborg2015}. The islands exhibit a characteristic hexagonal pattern reflecting a moir\'e structure which results from the lattice mismatch between the Ag(111) surface and the \mos\ (\Figref{fig1}b). Areas with large apparent height correspond to domains, where the S atoms sit on top of Ag atoms, whereas the lower areas represent two different hollow sites (fcc or hcp stacking) of the S atoms on the Ag lattice. 
The most abundant moir\'e periodicity amounts to $\sim 3.3\pm 0.1$\,nm. This value is similar to the one observed for \mos\ on Au(111) \cite{Gronborg2015, Sorensen2014, Bruix2016,Bana2018}.  
Given the very comparable lattice constants of Au (4.08\, \AA) and Ag (4.09\, \AA), a locking into a similar superstructure at the metal--\mos\ interface is not surprising. However, occasionally, we also observe moir\'e patterns with $3.6\pm 0.1$\,nm and $3.0\pm 0.1$\,nm lattice constants and different angles between the \mos\ and Ag(111) lattice. This indicates shallow energetic minima of the lattice orientations.
Atomically resolved STM images (\Figref{fig1}c) reveal the expected S--S distance of 3.15\,\AA\ in the top layer \cite{Wakabayashi1975, Bronsema1986,Schumacher1993,Helveg2000}. 

\begin{figure*}
\includegraphics[width=16cm]{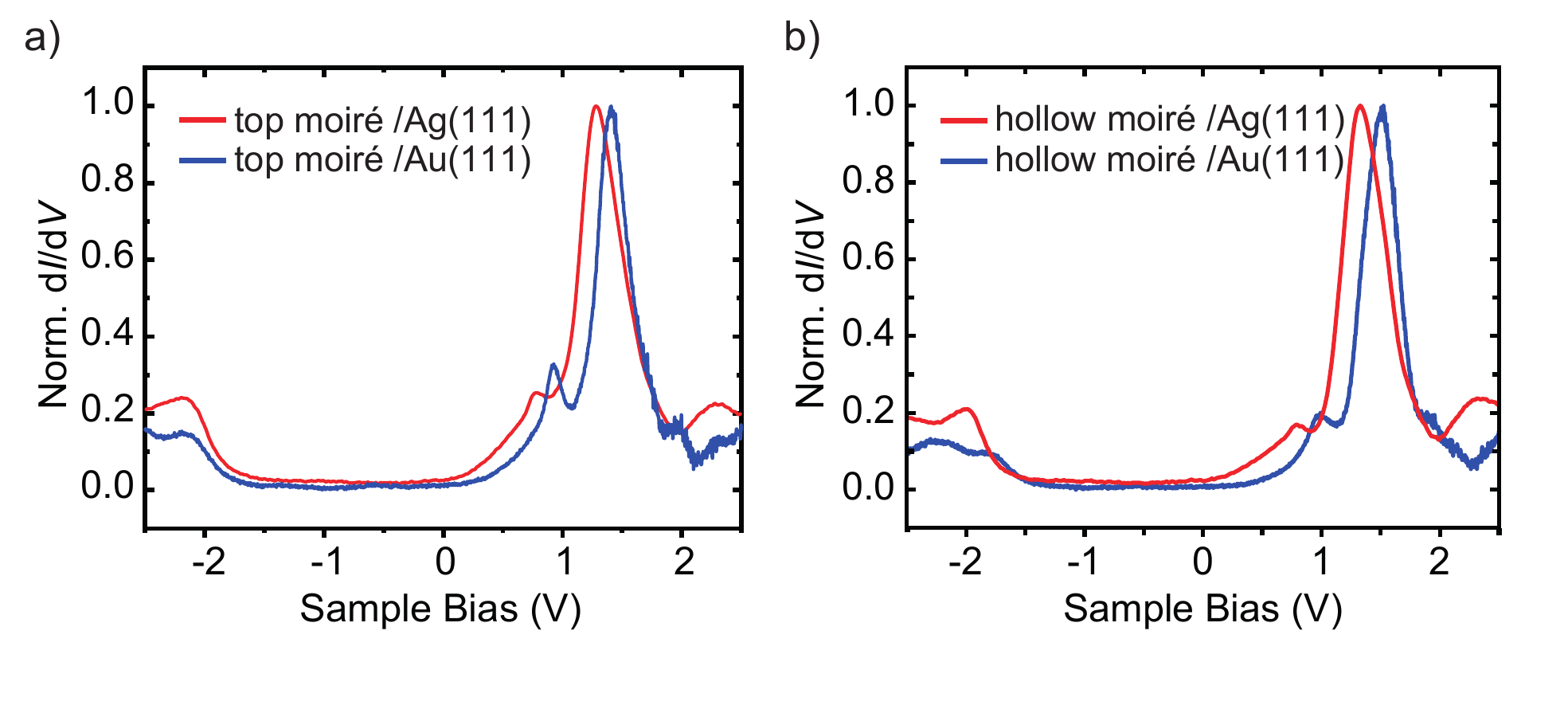}
\caption{Constant-height \didv spectra recorded (a) on top and (b) on hollow site of the moir\'e structure of \mos\ on Ag(111) (red curves) and on Au(111) (blue curves). Feedback opened at V = 2.5\,V, I = 0.5\,nA, V$_\mathrm{mod}=10$\,mV (all spectra, except for hollow site on Au(111): V$_\mathrm{mod}=5$\,mV).  }
\label{fig2}
\end{figure*}

For an efficient decoupling of a molecule from the substrate, the interlayer must provide an electronic band gap. As the moir\'e pattern bears a topographic and an electronic modulation \cite{Krane2018a}, we investigate the differential conductance (\didv) spectra on different locations (\Figref{fig1}d). We first examine the spectrum on the top site of the moir\'e structure. We observe a gap in the density of states, which is flanked by an onset of conductance at $\sim -1.55$\,V and $\sim +0.05\,$V (marked by dashed line labelled VBM/CBM, which have been determined from a logarithmically scaled plot). Additionally, there are pronounced peaks at $\sim 0.77$\,V and $\sim 1.28$\,V. First, we note that the observed band gap is significantly smaller than the 2.8\,eV-band-gap of a single layer free-standing \mos\ \cite{Cheiwchanchamnangij2012, Qiu2015}. This indicates a strong hybridization of the electronic states of the \mos\ layer and the Ag substrate. Second, we note that the spectral features are similar to those observed for single-layer \mos\ on Au(111) \cite{Krane2016, Bruix2016, Krane2018a}. For direct comparison, we plot the spectra on the top sites of the \mos\ moir\'e on Au(111) and Ag(111) in \Figref{fig2}a. At negative bias voltage, the onsets of conductance are essentially the same, while the features at positive bias voltage appear $\sim 140$\,mV  closer to the Fermi level on Ag(111) than on Au(111). 

Before discussing the differences between the layers on Au(111) and Ag(111), we investigate the effect of the different stacking at the interface on the electronic properties. The spectrum on a hollow site on Ag(111) shows a shift of the features at negative bias voltage by about $\sim 130$\,mV towards the Fermi level ($E_\mathrm{F}$), whereas the peaks at positive bias undergo a much smaller shift ($\sim 50$\, mV) away from $E_\mathrm{F}$ (\Figref{fig1}d). On Au(111), there are also  variations between hollow and top sites, with the strongest shift at negative bias voltage (\Figref{fig2}).

To understand the differences between the substrates and local sites, we first discuss the origin of the spectroscopic features. Based on the similarity of the spectral shapes on Au(111) and Ag(111), we tentatively assign the strong peaks at $\sim 0.8$\,V (labeled as $\Gamma_1$) and $\sim 1.3$\,V (labeled as $\Gamma_2$) (values averaged over the different moir\'e sites) to bands at the $\bar{\Gamma}$ point \cite{Krane2018a}. More precisely, the peak at $\Gamma_2$ has been assigned to bands at $\Gamma$, which are also present in free-standing \mos, but are broadened due to hybridization with the substrate. The peak at $\Gamma_1$ has been observed in tunneling spectra on \mos\ on Au(111), but has not been found in calculations. It has been interpreted as a hybrid metal-\mos\ or interface state \cite{Krane2018a}. The conduction band minimum, which is expected to lie at the $\bar{K}$ point for quasi free-standing as well as metal-supported single-layer \mos\ \cite{Mak2010, Splendiani2010,Miwa2014,Bruix2016} is hardly visible in the tunneling spectra due to the rapid decay of the tunneling constant with $k_{\parallel}$  \cite{Zhang2015, Krane2018a}. The same applies to the valence band maximum, such that the strongest feature in the tunneling spectra at -2\,V arises from bands close to the $\bar{Q}$ point \cite{Krane2018a}. 

Comparison of spectra on the moir\'e hollow sites suggest a rigid shift of the conduction bands between the \mos\ bands on Ag and Au. In a very simple interpretation, this agrees with the lower work function of Ag than of Au. 
A down-shift of the conduction band structure by $\sim 280$\, meV has been observed by photoemission of WS$_2$ on Au(111) and Ag(111) \cite{Dendzik2017}. Angle-resolved measurements further showed that the shift also included band distortions, such that bands at $Q$ were crossing $E_\mathrm{F}$ (instead of at $K$). The band distortion was explained by hybridization of the WS$_2$ bands with the Ag substrate \cite{Dendzik2017}. As our \didv signal is not $k_\parallel$-sensitive, we would not be able to detect band distortions in the \mos --Ag system. However, the clear shift of the states at $\Gamma$ can be easily understood by hybridization of S-derived states of mainly out-of-plane character with Ag states in analogy to Ref. \cite{Bruix2016}. 

In the occupied states, the bands on the hollow site follow the same trend of a down-shift, suggesting that the states near $\bar Q$ are equally affected by hybridization with Ag states \cite{Dendzik2017}. In contrast, the tunneling spectra on the top sites, seem to coincide for Au and Ag substrate. We also note that the tunneling conductance close to the $\bar Q$ point is the most sensitive to the precise location on the moir\'e pattern. Hence, we suggest that this site is most strongly affected by screening effects, which may vary on the different substrates \cite{Roesner2016} and partially compensate for hybridization effects.

\subsection{Electronic properties of TCNQ molecules on \mos\ on Ag(111)}

\begin{figure}
\includegraphics[width=\columnwidth]{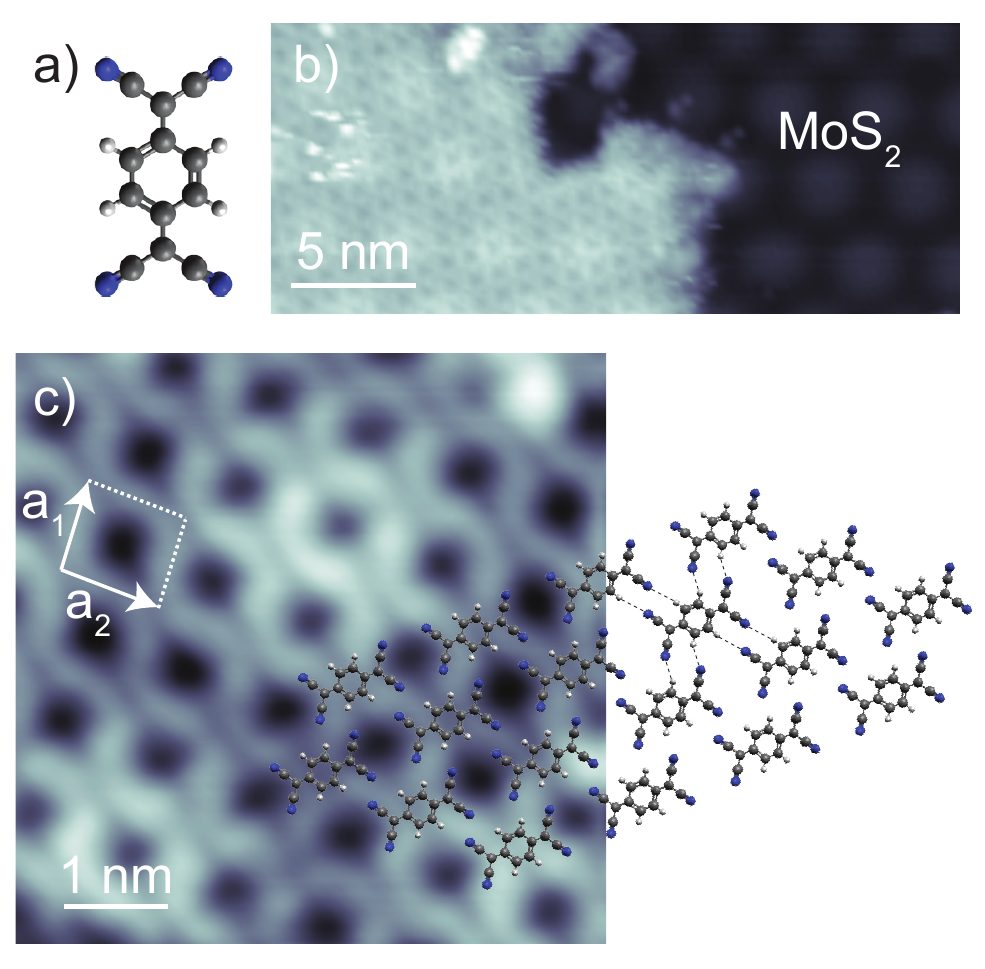}
\caption{ a) Stick-and-ball model of TCNQ. Gray, blue and white spheres represent C, N and H atoms, respectively. 
b) STM topography of a TCNQ molecular island on \mos/Ag(111) recorded at V = 1\,V, I = 10\,pA.  
c) STM topography of a TCNQ island on MoS$_2$/Ag(111) recorded at V = 0.8\,V, I = 200\,pA, with superimposed molecular models suggesting intermolecular  dipole-dipole interactions (dashed lines). White arrows represent the unit cell of the self-organized TCNQ domain with lattice vectors $a_1$ = (0.9 $\pm$0.10)\,nm and $a_2$ = (1.0 $\pm$0.10)\,nm and the angle between them of (96$\pm$2)$^{\circ}$. }
\label{fig3}
\end{figure}

\begin{figure}
\includegraphics[width=\columnwidth]{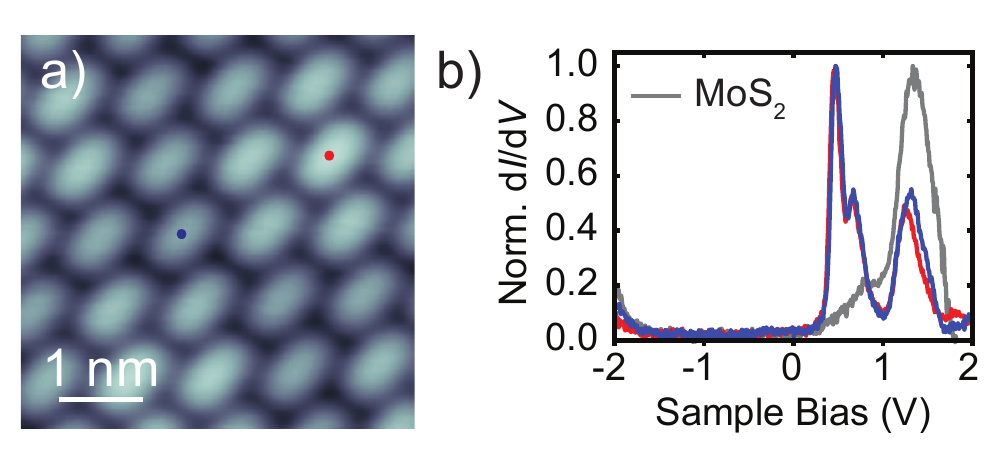}
\caption{a)  STM topography of a self-assembled TCNQ island on \mos/Ag(111), recorded at V = 0.2\,V, I = 20\,pA.
b) \didv spectra acquired on TCNQ molecules within the island in (a), with the precise location marked by colored dots. The gray spectrum was recorded on bare \mos\ layer for reference. Feedback opened at V = 2\,V, I = 100\,pA, with $V_\mathrm{mod} =20$\,mV.} 
\label{fig4}
\end{figure}

Deposition of TCNQ molecules (structure shown in \Figref{fig3}a) on the sample held at 230\,K leads to large densely packed molecular islands on the \mos\ areas (\Figref{fig3}b). The large size and high degree of order of these islands reflects a low diffusion barrier on the \mos\ substrate. 
The moir\'e pattern of \mos\ remains intact and visible through the molecular monolayer. High-resolution STM images recorded at  0.8\,V (\Figref{fig3}c) allow to resolve the individual molecules and their arrangement. Each TCNQ molecule appears with back-to-back double U-shapes separated by a nodal plane. As will be discussed later, and based on previous work on TCNQ \cite{Torrente2008, Garnica2013}, this appearance can be associated to the spatial distribution of the lowest unoccupied molecular orbital (LUMO). The molecular arrangement can be described by the lattice vectors $a_1=0.9\pm 0.1$\,nm, $a_2=1.0 \pm$0.1\,nm and angle (96$\pm$2)$\, ^{\circ}$ (see model in \Figref{fig3}c). This structure is stabilized by dipole-dipole interactions between the cyano endgroups and the quinone center of neighboring molecules.  This assembly is very similar to typical self-assembled TCNQ islands on weakly interacting substrates \cite{Torrente2008, Barja2010, Garnica2013, Park2014, Pham2019}. When measured at lower bias voltage (e.g., at V = 0.2\,V in \Figref{fig4}a), the molecules appear with featureless elliptical shape, reflecting only the topographic extent of the molecules.

\begin{figure*}
\includegraphics[width=16cm]{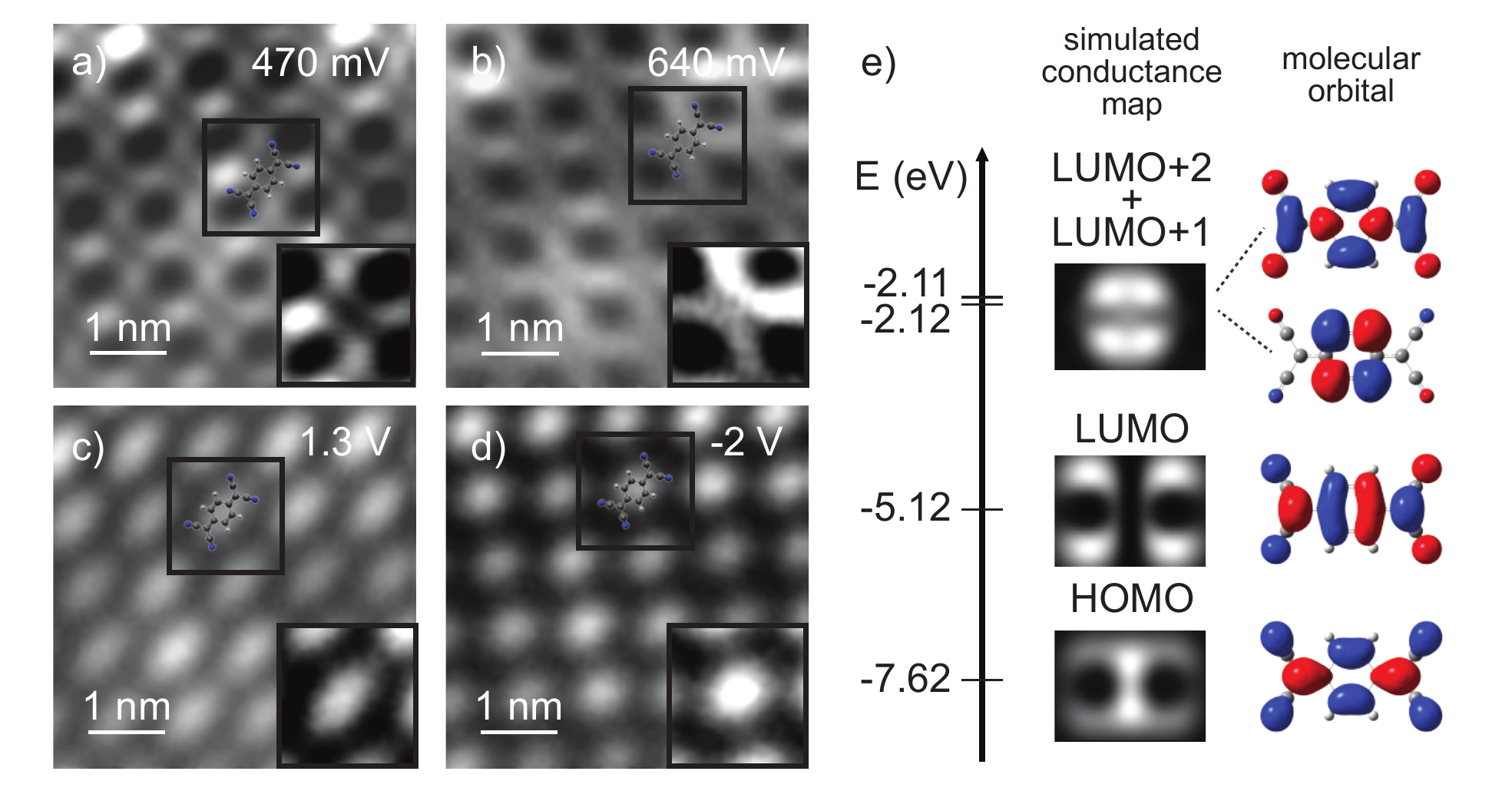}
\caption{a-d) Constant-height \didv maps of a TCNQ island on \mos\ recorded at the resonance energies derived in \Figref{fig4}b. Feedback opened at (a-c)  V = 2\,V, I = 100\,pA and (d) V = -2\,V, I = 30\,pA on the center of molecule with $V_\mathrm{mod} =20$\,mV. Zoom with increased contrast on one molecule are shown as inset for each map
e) Energy-level diagram of TCNQ determined from gas-phase DFT calculations (left). The isosurfaces of the frontier molecular orbitals are shown on the right. These have been used to calculate the tunneling matrix element $M_\mathrm{ts}$ with an s-wave tip at a tip--molecule distance of 7.5\,\AA, workfunction of 5\,eV. The map of the spatial distribution of $\left|M_\mathrm{ts}\right|^2$ is shown in the middle panel.	
}
\label{fig5}
\end{figure*}

The strong bias-voltage dependence of the TCNQ molecules on the \mos\ layer promises energetically well separated molecular states. To investigate these properties in more detail, we recorded \didv spectra on top of the molecules (\Figref{fig4}b). These show two main resonances at $\sim 0.47$\,V and $\sim 0.64$\,V. Another peak at $\sim 1.3$\,V matches the $\Gamma$ resonance of the bare \mos\ layer. At negative bias voltage, we observe an onset of conductance at $\sim -1.8$\,V. The \didv spectra thus show that the STM image in \Figref{fig4}a was recorded within the energy gap of the molecule, which explains the featureless shape. 
In order to determine the origin of each of the resonances, we recorded constant-height \didv maps at their corresponding energies (\Figref{fig5}).

For the first resonance at positive bias voltage (470\,mV, \Figref{fig5}a), we observe the same double U-shape, separated by a nodal plane, which we used in \Figref{fig3} for the identification of the molecular arrangement. The \didv map at 640\,mV exhibits the same shape, suggesting the same orbital as its origin. At 1.3\,V, the molecules do not show any characteristic feature (\Figref{fig5}c). Finally, \Figref{fig5}d presents a conductance map at -2\,V associated with the onset of conductance observed at negative bias voltage for spectra on the molecule. Here, the \didv signal is rather blurred, but we remark that it is more localized in the center of the molecule as compared to the elliptical shape in \Figref{fig5}c. 

For the identification of molecular orbitals, it is often sufficient to compare the \didv maps with the shape of the gas-phase molecular orbitals. Using this method, the U-shaped features have previously been associated to the LUMO of TCNQ \cite{Torrente2008, Garnica2013, Pham2019}. Here, we corroborate this assignment by simulating constant-height \didv maps of a free, flat-lying molecule. We first calculated the gas-phase electronic structure using density-functional-theory calculations with the B3PW91 functional and the 6-31g(d,p) basis set as implemented in the GAUSSIAN09 package \cite{Gaussian}. The isodensity contour plots of the highest occupied molecular orbital (HOMO) and some of the lowest unoccupied orbitals are shown in \Figref{fig5}e, right panel. The HOMO/LUMO can be unambiguously distinguished by the absence/presence of a nodal plane at the center of the quinone backbone. For direct comparison with the \didv maps, we calculate the tunneling matrix element between an s-wave tip and the spatially-resolved molecular wavefunction across the molecule \cite{Bardeen1961}. The maps of the square of the tunneling matrix element are depicted in \Figref{fig5}e next to the corresponding molecular orbitals.  Because the LUMO+1 and LUMO+2 are quasi degenerate, we used the sum of their wave functions for the calculations of the tunneling matrix elements. As expected, the nodal planes of the molecular orbitals dominate the simulated \didv maps and can be taken as a robust signature for molecular orbital identification. Additionally, the simulated maps reveal that \didv intensity is not directly proportional to the isosurface density. For instance, there is hardly any intensity within the U shapes of the TCNQ LUMO, and the HOMO is mainly localized at the very center of the quinone moiety. We note that the simulated maps were obtained at a tip-molecule distance (center of the s-wave tip to center of the molecule) of 7.5\,\AA. This value was chosen because it represents reasonable tunneling conditions in experiment. However, variation of the tip height by ($\pm$2\,\AA) does not have any influence on the observation of the main features within the map (i.e., nodal planes, or intensity maxima) \cite{Reecht2020}.

Comparison with the experimental constant-height \didv maps, now allows for an unambiguous identification of the origin of the molecular resonances. As suggested previously, the resonance at 0.47\,V can be derived from the LUMO with the double U-shape being in very good agreement to the calculations of the tunneling matrix element. The very same signatures in the conductance map at 0.64\,V suggest that this resonance stems from the LUMO as well. The DFT calculations show that the LUMO is non-degenerate. Hence, we can exclude a substrate-induced lifting of the degeneracy. The energy difference of only 170\,meV between the two resonances lies within the typical vibrational energies of organic molecules and may, thus, be indicative of a vibronic peak. We will elucidate on this point further below. 

The \didv map at 1.3\,V essentially shows the same elliptical shapes of the molecules as the STM image recorded in the electronic gap (\Figref{fig4}a). Our DFT calculations suggest that the next higher unoccupied orbitals lie 3\,eV above the LUMO and show a pattern of nodal planes that are absent in the experiment.
Additionally, given the similar energy with the \mos\ bands, this resonance is probably not associated to the molecular layer, but to direct tunneling into the \mos\ states. 

The assignment of the orbital origin at negative bias voltage bears some intricacies, because the experimental map lacks characteristic nodal planes. The reduced spatial resolution is most probably caused by the overlap with density of states of the substrate as we are approaching the onset of the valence band of \mos. One may suggest that the stronger localization of \didv intensity toward the quinone center is in agreement with the large tunneling matrix element of the HOMO at the molecule's center. This assignment may be enforced by the coincidence of the observed molecular energy gap of TCNQ with the DFT-derived gap. However, DFT is known to underestimate HOMO--LUMO gaps. Though this effect may be compensated by the screening properties of the substrate, we refrain from a definite assignment. In any case, our data clearly shows that the HOMO is at or within the conduction band of \mos.

By comparison with simulations, we thus arrive at a clear identification of the energy level alignment. Most notably, we find that the LUMO-derived resonance lies close to, but above the Fermi level of the substrate, whereas the HOMO is far below. This leaves the molecule in a neutral state with a negligible amount of charge transfer, despite of the electron accepting character of TCNQ. Nonetheless, its electron affinity of $\sim$3.4\,eV \cite{Milian2004,Zhu2015} is consistent with the LUMO alignment just above $E_\mathrm{F}$ when considering the work function of \mos/Ag(111) of 4.7\,eV \cite{Zhong2016}. We found small shifts of the LUMO onsets by at most 50\,mV between the spectra of TCNQ molecules lying at the top or hollow sites of the moir\'e structure of \mos. These shifts correspond to the moir\'e-induced shifts in unoccupied states of the \mos\ layer and thus only reflect the different screening properties from the substrate. In turn, we do not observe any modification of the electronic structure of \mos. This indicates weak interactions of the molecules all along the \mos\ layer. 

Importantly, the 470-meV resonance has a rather narrow width of $\sim 100$\,meV. This is significantly smaller than typically observed on metal surfaces, where strong hybridization effects lead to widths of the order of $\sim500$\,meV \cite{Torrente2008,Park2014}. The narrow width thus reflects that \mos\ acts as a decoupling layer from the metal substrate. However, this resonance width is broader than has been observed for the HOMO resonance of other organic molecules on \mos\ on Au(111) \cite{Krane2018, Reecht2019,Reecht2020}.  In contrast to those cases, where the HOMO lay well inside the electronic gap of \mos, the LUMO of TCNQ is located right at the onset of the conduction band. This provides relaxation pathways for electrons tunneling into the LUMO, though still significantly less than on the bare metal.

\subsection{Vibronic excitations of TCNQ on \mos\ on Ag(111)}

\begin{figure*}[h]
\includegraphics[width=16cm]{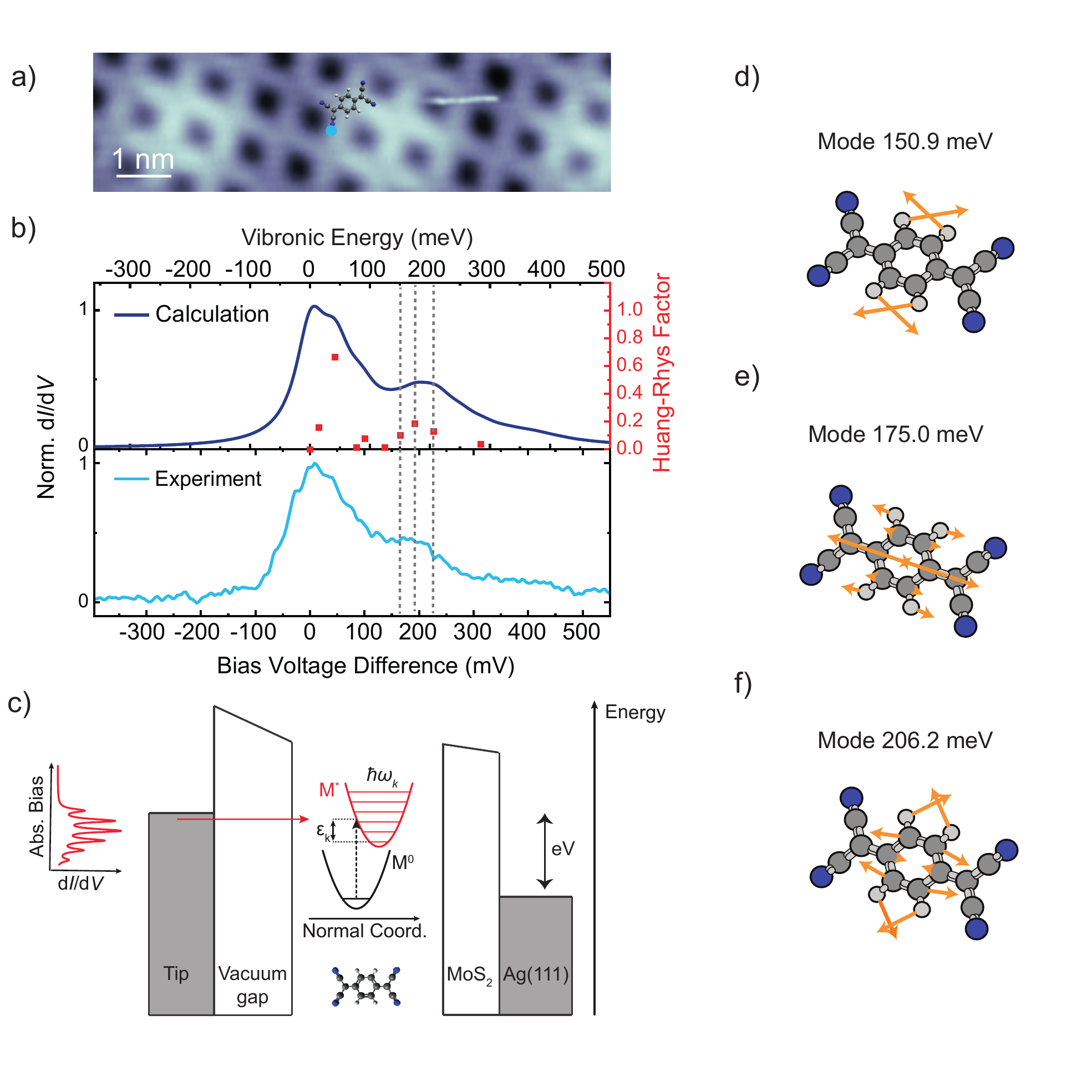}
\caption{a) STM topography of a TCNQ island recorded at V = 1\,V, I = 10\,pA. b) Simulated (top panel) and experimental (bottom panel) \didv spectra at the position indicated by the blue dot in a) with feedback opened at V = 2\,V, I = 100\,pA, with $V_\mathrm{mod} =10$\,mV. The simulated spectrum is obtained from DFT calculations for all the vibrational mode of the TCNQ$^{-}$ molecule with a Huang-Rhys factor higher than 0.01 (dots associated with the right axis). A Lorentzian peak of 60\,meV broadening is applied to all of these modes. 
c) Schematic representation of electron transport through a TCNQ molecule adsorbed on \mos/Ag(111): singly charged TCNQ$^{-}$ is formed upon injecting an electron into a vibronic state of an unoccupied molecular electronic level.
d-f) Visualization of the vibrational modes contributing to the satellite peak. The orange arrows represent the displacement of the atoms involved in these vibrations.
}
\label{fig6}
\end{figure*}

Having shown that the resonances at 470\,mV and 640\,mV originate both from the LUMO of TCNQ, we now turn to their more detailed analysis. A close-up view of the spectral range with these peaks is shown in the bottom panel of \Figref{fig6}b with the LUMO-derived peak at 470\,mV shifted to zero energy and its peak height being normalized. The satellite structure is reminiscent of vibronic sidebands, which occur due to the simultaneous excitation of a vibrational mode upon charging \cite{Qiu2004, Pradhan2005,Nazin2005,Frederiksen2008,Matino2011,Schulz2013,Wickenburg2016}. The sidepeaks should thus obey the same symmetry as the parent orbital state \cite{Huan2011,Schwarz2015,Mehler2018}. In the simplest case, these excitations can be described within the Franck-Condon model (see sketch in \Figref{fig6}c). When probing the LUMO in tunneling spectroscopy, the molecule is transiently negatively charged. Within Born-Oppenheimer approximation, this process is described by a vertical transition in the energy level diagram from the ground state $M^{0}$ to the excited state $M^{*}$. Upon charging, the molecule undergoes a geometric distortion, captured by the shift of the potential energy curve of the excited state. Vertical transitions allow for probing many vibronic states, with the intensities given by a Poisson distribution $I_{kn} = e^{-S_k} \frac{S_k^n}{n!}$, with $S_k$ being the Huang-Rhys factor of the vibrational mode $k$ and $n$ its harmonics. The Huang-Rhys factor is determined by the relaxation energy $\epsilon_k$ of a vibrational mode when charging the molecule as $S_k = \frac{\epsilon_k}{\hslash \omega_k}$. From the DFT calculations of the TCNQ molecule, we determine all vibrational eigenmodes in the negatively charged state and also derive the Huang-Rhys factors $S_k$ \cite{Krane2018}. The latter is plotted in the upper panel of \Figref{fig6}b (dots, right axis).  
Applying to each of the vibronic states a Lorentzian peak with a full width at half maximum of 60\,meV and intensity proportional to the Poisson distribution, as described above, leads to a simulated Franck-Condon spectrum in the upper panel of \Figref{fig6}b.
This spectrum closely resembles the experimental one and, therefore, nicely reflects the nature of the satellite structure. We note that the bias voltage axis (bottom panel) is scaled by $10\%$ compared to the energy axis (top panel) to account for the voltage drop across the \mos\ layer \cite{Krane2019}. We now realize that the peak at $\sim 640$\,meV consists of three vibrational modes (at 151\,meV, 175\,meV and 206\,mV) exhibiting a large Huang-Rhys factor. These modes correspond to in-plane breathing modes of TCNQ (see schemes in \Figref{fig6}d-f), which are particularly sensitive to charging. Additionally, a mode at 40\,meV has a large Huang-Rhys factor. The excitation of this mode is not energetically well separated from the elastic onset of the LUMO in experiment. However, this mode contributes to an asymmetric lineshape, which can be realized by comparing the low-energy flank to the high-energy fall-off of the first resonance. The low-energy side can be fitted by a Voigt profile and suggests a lifetime broadening of $55\pm 15$\,meV. This is, however, insufficient for a peak separation from the 40-meV mode. 

We further note that the experimental spectrum was taken on a cyano group, where no nodal planes exist in the LUMO, as their presence may lead to vibration-assisted tunneling in addition to the bare Franck-Condon excitation \cite{Reecht2020}.

\section{Conclusions}
We have shown that a single layer of \mos\ may act as a decoupling layer for molecules from the underlying metal surface, if the molecular resonances lie within the semiconducting band gap of \mos. \mos\ on Au(111) and Ag(111) exhibit very similar gap structures, but are shifted in energy according to the different work functions of the metal. Though this is not the only reason for the band modifications \cite{Dendzik2017}, we suggest that such considerations may help when searching for appropriate decoupling layers for specific molecules. 
We have challenged the decoupling properties of \mos/Ag(111) for TCNQ molecules. These exhibit their LUMO resonance just at the conduction band onset of \mos, whereas the HOMO lies within the valence band.
Hence, the HOMO is not decoupled from the substrate, and also the LUMO suffers considerable lifetime broadening as compared to resonances, which would be well separated from the onsets of the \mos\ bands. The lifetime 
broadening of $55\pm 15$\,meV can be translated into a lifetime of $\sim $6\,fs of the excited state. This is almost one order of magnitude longer than on the bare metal surface, where the hot electron vanishes into the bulk on ultra-fast timescales, but an order of magnitude shorter than for molecular resonances well separated from the band onsets \cite{Krane2018, Reecht2019, Reecht2020}. Yet, the increase in the lifetime of the excited state allowed us to resolve vibronic states of the transiently negatively charged TCNQ molecule albeit only up to $\sim 200$\,meV above the LUMO resonance, where contributions of \mos\ bands at $\Gamma$ become strong. Our simulations reproduce the experimental satellite structure of the LUMO very well, although the experimental width prevented us from resolving the individual modes.

\begin{acknowledgements}
We acknowledge discussions with S. Trishin and J. R. Simon. A. Yousofnejad acknowledges a scholarship from the Claussen-Simon Stiftung. 
This work was supported by the Deutsche Forschungsgemeinschaft (DFG) - project number 182087777 - SFB 951 (A14), and project number 328545488 - TRR 227 (B05).
\end{acknowledgements}

\bibliographystyle{apsrev4-1}
%

\end{document}